\documentclass[%
reprint,
amsmath,amssymb,
aps,
]{revtex4-2}

\usepackage{graphicx}
\usepackage[colorlinks=true,linkcolor=blue,anchorcolor=blue, citecolor=blue,urlcolor=blue]{hyperref}

\begin{document}
	\preprint{APS/123-QED}

\title{Proximity effect and inverse proximity effect in a topological-insulator/iron-based-superconductor heterostructure}
\author{Qi-Guang Zhu$^{1}$}
\author{Tao Zhou$^{1,2}$}%
\email{Corresponding author: tzhou@scnu.edu.cn}
\affiliation{$^1$Guangdong Provincial Key Laboratory of Quantum Engineering and Quantum Materials, School of Physics and Telecommunication Engineering, South China Normal University, Guangzhou 510006, China\\
$^2$Guangdong-Hong Kong Joint Laboratory of Quantum Matter, Frontier Research Institute for Physics, South China Normal University, Guangzhou 510006, China
}
\begin{abstract}
We theoretically study the proximity effect and the inverse proximity effect in a topological insulator/iron-based superconductor heterostructure based on the microscopic model. The superconducting order parameter is self-consistently calculated. Its magnitude decreases when the coupling of these two systems increases. The induced pairing order parameter exhibits negative and positive values, while the negative pairing near the $\Gamma=(0,0)$ point is dominant.
This parameter has twofold symmetry and includes an $s$-wave component and a $d$-wave component. The magnitude of the induced pairing order parameter has a maximal value at the coupling strength $t_p=0.3\approx 0.06$ eV.
The spectral function and the local density of states are calculated and may be used to probe the proximity effect. We also discuss
the feedback of the topological insulator to the iron-based superconductor layer.
The normal-state Fermi surface is distorted by the coupling, and additional Fermi pockets are induced. An effective spin-orbit interaction term is induced. In the superconducting state, the previous fourfold symmetry of the order parameter is broken, and a $d$-wave component pairing term is also induced.
Our main results can be well understood by analyzing the Fermi surfaces of the original systems.
\end{abstract}
\maketitle

\section{introduction}

The heterostructure coupling a three-dimensional topological insulator (3DTI) material and a superconducting material has attracted great interest in the field of condensed matter physics. Theoretically, if superconducting order is proximity induced in the surface
of the 3DTI, then the system is predicted to become an effective $p+ip$ superconductor~\cite{PhysRevLett.100.096407}. Majorana zero modes can be realized in this platform; thus, this system has potential for application in the field of topological quantum computation~\cite{RevModPhys.80.1083}. Experimentally, this hybrid system was successfully realized, and signatures of Majorana zero modes were indeed observed~\cite{Wang2012,PhysRevLett.114.017001,PhysRevLett.116.257003,PhysRevLett.125.136802}.

The original proposed heterostructure system includes a conventional $s$-wave superconductor and a 3DTI~\cite{PhysRevLett.100.096407}.
For conventional superconductors, the superconducting transition temperature is usually very low, and the pairing gap is rather small (approximately 1 meV). Then, in the presence of a vortex, the excited low-energy states are extremely small.
Experimentally differentiating the Majorana zero modes from the excited low-energy states is rather challenging.
Thus, employing a high-T$_c$ superconducting material to realize a topological superconductor and Majorana zero modes is natural.
Many efforts have been made to realize an effective topological superconductor with a high-T$_c$ superconductor platform.
Hybrid systems with a 3DTI and a cuprate-based high-T$_c$ superconductor were experimentally realized~\cite{Zareapour2012,wang2013fully,PhysRevB.101.220503,PhysRevLett.113.067003,PhysRevB.90.085128}. However, the results seem to be controversial. A large proximity-induced gap of approximately 10 meV at the surface of the 3DTI was reported by several groups~\cite{Zareapour2012,wang2013fully,PhysRevB.101.220503}, while this result was challenged by other groups~\cite{PhysRevLett.113.067003,PhysRevB.90.085128}. The heterostructure with a 3DTI and a member of the family of iron-based high-T$_c$ superconductors (FeSCs) was also realized by many groups, and signatures of the proximity-induced superconducting gap were revealed~\cite{He2014,PhysRevB.97.224504,doi:10.1126/sciadv.aat1084,doi:10.1021/acs.nanolett.1c01703,ZHANG2022163396,zhang2022anomalous,PhysRevB.100.241402,doi:10.1073/pnas.1914534117}.

On the theoretical side, several 3DTI/superconductor coupled systems have been studied~\cite{PhysRevLett.100.096407,PhysRevLett.104.067001,PhysRevB.87.220506,PhysRevB.91.235143,PhysRevB.93.035140,PhysRevB.96.064518,PhysRevB.99.094505,PhysRevB.101.054512}. Generally, there are two theoretical methods to study the proximity effect in a heterostructure system. One is to phenomenologically treat the proximity effect by putting the superconducting pairing term directly in the Hamiltonian of the 3DTI material~\cite{PhysRevLett.100.096407,PhysRevLett.104.067001}. In this way, the model is greatly simplified, and the topological features are expected to be qualitatively described. An alternative method is to treat the proximity effect in a microscopic way~\cite{PhysRevB.87.220506,PhysRevB.91.235143,PhysRevB.93.035140,PhysRevB.96.064518,PhysRevB.99.094505,PhysRevB.101.054512,PhysRevB.81.241310,PhysRevB.103.024517}, namely, the whole Hamiltonian includes the 3DTI part, the superconductor part, and their coupling term.
The effective pairing term in the 3DTI system is induced by the coupling term. Since the mixing of the band structures of different materials is fully taken into account by the latter method, the numerical results may be more realistic. Previously, based on the latter method, a 3DTI/cuprate-based superconductor coupled system was theoretically studied. The results indicated that the proximity-induced pairing gap in the surface of the 3DTI becomes an $s$-wave gap~\cite{PhysRevB.91.235143,PhysRevB.93.035140}. This theoretical result is consistent with experiments~\cite{wang2013fully}. Another advantage of the latter method is that it can provide some useful information about the inverse proximity effect, namely, the possible feedback of the 3DTI system to the superconductor layer~\cite{PhysRevB.96.064518,PhysRevB.99.094505,PhysRevB.101.054512}.

Although the 3DTI/FeSC coupled system has recently attracted much experimental interest~\cite{He2014,doi:10.1021/acs.nanolett.1c01703,ZHANG2022163396,zhang2022anomalous,PhysRevB.97.224504,PhysRevB.100.241402,doi:10.1073/pnas.1914534117,doi:10.1126/sciadv.aat1084},
this coupled system has been less explored theoretically.
In particular, as far as we know, no theoretical studies have explored the coupled system with an FeSC material microscopically.
Actually, this issue is of importance and thus is necessary to study. First, in the normal state of the FeSC material, there are several energy bands crossing the Fermi energy. Then, the Fermi surface includes both hole pockets and electron pockets~\cite{RevModPhys.83.1589}.
Each energy band and Fermi pocket may couple with the band structure of the 3DTI. As a result, for the FeSC-based heterostructure, the band mixing effect may be more complex and important in affecting the physical properties of the system. Second, for the 3DTI system, the Fermi surface is a point, coming from the topological edge states. The sizes of the normal-state Fermi pockets of cuprate and conventional superconductors are usually rather large, while for iron-based superconductors, the normal-state Fermi pockets may be very small. Since the proximity effect may be strengthened if the normal-state Fermi surfaces of the two compounds match, we expect that the FeSC material should be an ideal candidate to induce an effective superconducting pairing term in the 3DTI surface. Third,
as has been verified, usually, the induced pairing symmetry is not identical to the previous symmetry. This may be determined by the combined effect of band mixing and the pairing symmetry of the superconductor. Currently, the pairing symmetry of the FeSC material is well believed to be the unconventional $s_\pm$-pairing~\cite{RevModPhys.83.1589}.
Studying the proximity-induced pairing term and predicting the pairing symmetry induced in the surface of the 3DTI would be interesting.
Fourth, studying the inverse proximity effect and exploring the back action of the 3DTI system on the FeSC material is timely and of interest.
This may also help in investigating the mechanism and the intrinsic nontrivial topology of the FeSC material~\cite{RevModPhys.83.1589,10.1093/nsr/nwy142}.
Finally, experimentally confirming whether the pairing term is successfully proximity induced in the 3DTI layer is not easy. With a microscopic method, more details may be provided to identify the proximity-induced superconducting gap.

In this paper, motivated by the above considerations, we first adopt a microscopic model
considering a 3DTI material coupled with a single-layer FeSC material and self-consistently deal with the whole system. The proximity-induced pairing order parameter in the 3DTI system is explored. We propose that the induced pairing gap can be measured by an angle-resolved photoemission spectroscopy (ARPES) experiment~\cite{RevModPhys.75.473} or the scanning tunneling microscopy (STM) technique~\cite{RevModPhys.79.353}. Additionally, we study the inverse proximity effect and indicate how coupling affects the FeSC material.

The rest of the paper is organized as follows.
In Sec. II, we introduce
the model and present the relevant formalism. In Sec. III, we
report numerical calculations and discuss the obtained
results. Finally, we present a brief summary and closing remarks in Sec. IV.

\section{Model and formalism}
We start with a model including the 3DTI term, the single-layer FeSC term, and their coupling term, with
\begin{equation}
	H=H_{TI}+H_{SC}+H_{p}.
\end{equation}

$H_{TI}$ describes the 3DTI material~\cite{supp}. Considering the open boundary condition along the $z$-direction and periodic boundary condition in the $xy$-plane, $H_{TI}$ is expressed as~\cite{PhysRevB.85.195119}
\begin{align}  	
	H_{TI} ={} &\sum_{z,\boldsymbol{\mathbf{k}}}{}C_{z}^{\dagger}\big( \boldsymbol{\mathbf{k}} \big) \Big[ m\big( \boldsymbol{\mathbf{k}} \big) \sigma _0\tau _z+2A\sin k_x\sigma _x\tau _{x}+ \notag\big.\\
	&\phantom{=\;\;}\big.2A\sin k_y\sigma _y\tau _{x} \Big] C_z\big( \boldsymbol{\mathbf{k}} \big) -
	\Big[ \sum_{z,\boldsymbol{\mathbf{k}}}{}C_{z+1}^{\dagger}\big( \boldsymbol{\mathbf{k}} \big)\notag \\
	& \big( t\sigma _0\tau _z-iA\sigma _z\tau _{x} \big) C_z\big( \boldsymbol{\mathbf{k}} \big)
	+H.c. \Big],	
\end{align}
where $\mathrm{m}\left( \mathbf{k} \right) = m-2t\left( \cos k_x+\cos k_y \right)$.
$\sigma _i$ and $\tau_{i}$ ($i=x,y,z$) are Pauli matrices in the spin channel and orbital channel, respectively.
$\sigma _0$ is the identity matrix. The momentum ${\bf k}$ is defined in the $xy$-plane, with ${\bf k}=(k_x,k_y)$.
The vector $ C^{\dagger}_z({\bf k}) $ is expressed as $( c_{z1{\bf k}\uparrow}^{\dagger},c_{z2{\bf k}\uparrow}^{\dagger},c_{z1{\bf k}\downarrow}^{\dagger},c_{z2{\bf k}\downarrow}^{\dagger} )$, with $z=1,2,\cdots,N_z$.

$H_{SC}$ describes the single-layer FeSC material.
FeSC materials are mostly quasi-two-dimensional layered materials.
A two-dimensional Hamiltonian considering a single layer is expected to also qualitatively describe FeSC bulk materials~\cite{PhysRevLett.103.186402,PhysRevB.80.104503,PhysRevLett.101.087004,PhysRevB.78.144517,PhysRevB.77.220503}.
The superconductivity of an FeSC material is mainly contributed by the Fe$_2$As$_2$/Fe$_2$Se$_2$ layers.
The Fe ions form a square lattice. The As/Se ions are above or below the Fe-Fe plane. Thus, each unit cell generally contains two Fe ions and two As/Se ions.
Previously, a model considering two Fe ions in one unit cell was proposed~\cite{PhysRevLett.103.186402}, while
the authors of Ref.~\cite{PhysRevLett.104.089701} argued that the system has an internal symmetry, namely, the Fe-As-Fe distances and angles are the same for As ions above and
below the Fe-Fe planes. Moreover,
based on a first-principles band calculation, the low-energy bands are mainly contributed by the Fe-3d orbitals~\cite{PhysRevB.80.104503,PhysRevLett.101.087004,PhysRevB.78.144517,PhysRevB.77.220503}. The As ions play the role of mediating or enhancing the next-nearest-neighbor hopping~\cite{PhysRevLett.103.186402,PhysRevLett.104.089701,PhysRevB.80.104503,PhysRevB.77.220503}.
Considering only the Fe orbitals, due to the internal symmetry, the Hamiltonian can be described by a reduced unit cell with only one Fe ion. Previously, several
effective models considering a one-Fe-ion unit cell have been proposed~\cite{PhysRevB.80.104503,PhysRevLett.101.087004,PhysRevB.78.144517,PhysRevB.77.220503}.
Following Ref.~\cite{PhysRevB.77.220503}, we use the minimum two-band model with one unit cell containing only one Fe ion and
consider the $d_{xz}$ and $d_{yz}$ orbitals to describe the FeSC system~\cite{PhysRevB.77.220503},
\begin{eqnarray}
	H_{SC}=&\sum_{{\bf k}\tau}\varepsilon_{\tau{\bf k}}d^\dagger_{\tau{\bf k}\sigma}d_{\tau{\bf k}\sigma}+\nonumber\\ &\sum_{{\bf k}}(\varepsilon_{12{\bf k}}d^\dagger_{1{\bf k}\sigma}d_{2{\bf k}\sigma}+H.c.)+H_{\mathrm{int}}.
\end{eqnarray}
Here, $\tau=1,2$ and $\sigma=\uparrow,\downarrow$ are the orbital index and the spin index, respectively.
$\varepsilon_{\tau{\bf k}}$ is the intraorbital hopping term, with
$\varepsilon_{1{\bf k}}=-2t_1\cos k_x-2 t_2\cos k_x-4t_3\cos k_x \cos k_y$ and $\varepsilon_{2{\bf k}}=-2t_2\cos k_x-2 t_1\cos k_x-4t_3\cos k_x \cos k_y$.
$\varepsilon_{12{\bf k}}$ is the interorbital hopping term, with $\varepsilon_{12{\bf k}}=-4t_4\sin k_x\sin k_y$. $\Delta_{\bf k}$ is the superconducting pairing function.

We now discuss the interaction term $H_{\mathrm{int}}$ in Eq. (3). Generally, both interorbital and intraorbital interactions may exist and mediate the superconducting pairing for an FeSC material.
Previously, the interorbital pairing was studied, but the theoretical results were in stark contrast to existing experiments. Therefore, it was concluded that the interorbital pairing should be excluded for FeSC materials~\cite{PhysRevB.81.104504}.
On the other hand, the intraorbital next-nearest-neighbor coupling strength was proposed to be stronger than the nearest-neighbor coupling strength due to As-mediated hopping~\cite{PhysRevB.78.144514}. Moreover,
considering the intraorbital next-nearest-neighbor attractive interaction,
the pairing function from the self-consistent calculation has the form of $\cos k_x \cos k_y$~\cite{PhysRevB.81.104504,PhysRevB.78.144514,PhysRevLett.101.206404,PhysRevB.80.014523}, consistent with the $s_{\pm}$ pairing symmetry and previous experimental observations~\cite{PhysRevLett.101.057003,PhysRevB.78.144514,Nakayama_2009}. In the present work, as is usually done, we take into account an effective intraorbital next-nearest-neighbor attractive interaction as a pairing strength; then, the interaction term $H_{\mathrm{int}}$ is expressed as
\begin{equation}
	H_{\mathrm{int}}=-\sum_{\langle \bf{ij}\rangle^\prime\tau\sigma\sigma^\prime} V_\tau n_{\tau{\bf i}\sigma}n_{\tau{\bf j}\sigma^\prime},
\end{equation}
where $\langle \bf{ij}\rangle^\prime$ denotes the next-nearest-neighbor bond. $n_{\tau{\bf i}\sigma}=d^\dagger_{\tau{\bf i}\sigma}d_{\tau{\bf i}\sigma}$ is the on-site particle number operator ($\tau$ and $\sigma$ are the orbital and spin indices, respectively).
We now define the local mean-field pairing order parameter as $\Delta_{\tau{\bf ij}}=V_\tau\langle d_{\tau{\bf i}\uparrow}d_{\tau{\bf j}\downarrow}-d_{\tau{\bf i}\downarrow}d_{\tau{\bf j}\uparrow}\rangle$. Following Refs.~\cite{PhysRevB.78.144514,PhysRevLett.101.206404,PhysRevB.80.014523}, here, the mean-field order parameter is considered to have $s$-wave pairing symmetry and be independent of the bonds, $\Delta_{\tau{\bf ij}}\equiv \Delta_{\tau 0}$.
Moreover, due to the symmetry of the $d_{xz}$ and $d_{yz}$ orbitals of FeSC materials, both the pairing strength and the mean-field order parameter should be independent of the orbitals ($V_\tau\equiv V$ and $\Delta_{\tau 0}\equiv\Delta_0$)~\cite{PhysRevB.78.144514,PhysRevLett.101.206404,PhysRevB.80.014523}. Then, we can use a single value $\Delta_0$ to represent the superconductivity of the FeSC layer, with
\begin{equation}
	\Delta_0=\frac{V}{8N}\sum_{\tau\langle{\bf ij}\rangle^\prime}\langle d_{\tau{\bf i}\uparrow}d_{\tau{\bf j}\downarrow}-d_{\tau{\bf i}\downarrow}d_{\tau{\bf j}\uparrow}\rangle,
\end{equation}
with $N$ being the number of sites in the single FeSC layer.

At the mean-field level, by transforming $H_{\mathrm{int}}$ into momentum space, the Hamiltonian for the FeSC layer can be rewritten as
\begin{eqnarray}
	H_{SC}=&\sum_{{\bf k}\tau}\varepsilon_{\tau{\bf k}}d^\dagger_{\tau{\bf k}\sigma}d_{\tau{\bf k}\sigma}+\sum_{{\bf k}}(\varepsilon_{12{\bf k}}d^\dagger_{1{\bf k}\sigma}d_{2{\bf k}\sigma}+H.c.)\nonumber\\ &+\sum_{{\bf k}\tau}(\Delta_{\bf k}d^\dagger_{\tau{\bf k}\uparrow}d^\dagger_{\tau{-\bf k}\downarrow}+H.c.),
\end{eqnarray}
with $\Delta_{\bf k}=4\Delta_0\cos k_x\cos k_y$. It changes sign between electron and hole Fermi pockets, consistent with previous experimental and theoretical results~\cite{Nakayama_2009,PhysRevLett.101.057003,PhysRevB.78.144514}.

Performing the Fourier transformation on the self-consistent equation [Eq. (5)],
$\Delta_0$ can be calculated through the Hamiltonian in momentum space, with
\begin{equation}
	\Delta_0=\frac{V}{N}\sum_{\tau{\bf k}}\cos k_x\cos k_y \langle d_{\tau{\bf k}\uparrow}d_{\tau{-\bf k}\downarrow}\rangle.
\end{equation}

$H_p$ represents the coupling of the FeSC material and the surface of the 3DTI material, expressed as
\begin{equation}
	H_p=-\sum_{{\bf k}\tau\tau^\prime\sigma}(t_{p\tau\tau^\prime}d^\dagger_{\tau{\bf k}\sigma}c_{z\tau^\prime{\bf k}\sigma}+H.c.),
\end{equation}
with the interlayer hopping constant $t_{p\tau\tau\prime}$ being the coupling strength.
$z=1$ or $N_z$ represents the surface of the 3DTI material.

Considering one FeSC layer coupled with the 3DTI surface $(z=N_z)$, the whole Hamiltonian can be rewritten in matrix form as
$H=\sum_{\bf k}C^\dagger({\bf k})\hat{M}({\bf k})C({\bf k})$. The vector $C^\dagger({\bf k})$ is expressed as
\begin{eqnarray}
	C^\dagger({\bf k})=&\big( C^\dagger_1({\bf k}),C_1(-{\bf k}),\cdots,C^\dagger_{N_z}({\bf k}),C_{N_z}(-{\bf k}),\nonumber\\&D^\dagger({\bf k}),D(-{\bf k}) \big),
\end{eqnarray}
with $D^\dagger({\bf k})=( d_{1{\bf k}\uparrow}^{\dagger},d_{2{\bf k}\uparrow}^{\dagger},d_{1{\bf k}\downarrow}^{\dagger},d_{2{\bf k}\downarrow}^{\dagger} )$.
$\hat{M}({\bf k})$ is an $8(N_z+1)\times 8(N_z+1)$ matrix.

Diagonalizing the Hamiltonian matrix, the self-consistent equation for the superconducting order parameter $\Delta_0$ [Eq. (7)] can be rewritten as
\begin{align}  	 	
	\Delta _0  ={}& \frac{V}{N}\sum_{\mathbf{k},\tau,n}\cos k_x  \cos k_y  u_{r+\tau ,n}^{*}\left( \mathbf{k} \right) u_{r +\tau+6,n}\left( \mathbf{k} \right) f\left( E_n \right),
\end{align}
with $r=8N_z$. $u_{r,n}( \mathbf{k})$ and $E_n$ are the eigenvectors and eigenvalues of the Hamiltonian matrix. $f(x)$ is the Fermi distribution function.

We define an effective mean-field pairing order parameter at the surface of the 3DTI or the FeSC layer to study the proximity effect and the inverse proximity effect, expressed as
\begin{align}  	 	
	\Delta _{TI(Fe)}({\bf k})  =-\sum_{\tau,n}u_{s+\tau ,n}^{*}\left( \mathbf{k} \right) u_{s +\tau+6,n}\left( \mathbf{k} \right) f\left( E_n \right),
\end{align}
with $s=8(N_z-1)$ or $s=8N_z$ describing the intraorbital pairing order parameter at the surface of the 3DTI or the FeSC layer, respectively.
It is necessary to note that in principle, the interorbital pairing may also be proximity induced. We numerically verify that the magnitudes of the induced interorbital order parameters at the 3DTI surface are generally much smaller than those of the induced intraorbital order parameters (see the supplementary material~\cite{supp}).
Thus the induced interorbital mean-field pairing order parameters are not considered here.

The Green's function matrix can be obtained by diagonalizing the Hamiltonian, with the elements being expressed as
\begin{equation}  	
	G_{ij}({\bf k}, E) =\sum_n{\frac{u_{in}({\bf k})u_{jn}({\bf k})^{*}}{E-E_n+i\Gamma}}.
\end{equation}

Then, the $z$-dependent spectral function can be expressed as
\begin{align}  	
	A_z( \mathbf{k},E ) =-\frac{1}{\pi}\sum_{p=1}^4{\mathrm{Im}G_{m+p,m+p}}( \mathbf{k},E ),
\end{align}
with $m=8(z-1)$.

The local density of states (LDOS) at layer $z$ can be calculated through the spectral function, with
\begin{align}  	
	\rho _z( E ) =\sum_{\mathbf{k}
	}{A_z\left( \mathbf{k}
		,E \right)}.
\end{align}

In the original model describing the FeSC material ($H_{SC}$), there is no spin-orbit coupling term. Generally, the spin-orbit coupling term can be expressed as a spin-flipped hopping term with the form $d^\dagger_{{\bf k}\uparrow}d_{{\bf k}\downarrow}$. Here, due to the coupling with the 3DTI material, an effective spin-orbit coupling term may be induced in the superconducting layer. We define
a spin-flipped mean-field order parameter $\lambda({\bf k})=\langle d^\dagger_{{\bf k}\uparrow}d_{{\bf k}\downarrow}\rangle$ to describe this effective spin-orbit coupling term in the FeSC layer, with
\begin{align}  	 	
	\lambda({\bf k}) =\sum_{\tau,n}u_{r+\tau ,n}^{*}\left( \mathbf{k} \right) u_{r+\tau+2,n}\left( \mathbf{k} \right) f\left( E_n \right),
\end{align}
with $r=8N_z$.

In the present work, the coupling between the FeSC layer and the 3DTI material is considered an interlayer single particle hopping term, as indicated in Eq. (8). In principle, the signs and magnitudes of hopping constants $t_{p\tau\tau^\prime}$ may depend on the orbitals of the FeSC material and the 3DTI material. For the FeSC layer, $d_{xz}$ and $d_{yz}$
orbitals are considered~\cite{PhysRevB.77.220503}. The electronic distributions along the z-direction for these two orbitals are the same. As a
result, considering $t_{p1\tau^\prime} = t_{p2\tau^\prime}$ is reasonable.
In contrast, both the phases and magnitudes of the hopping constant for different 3DTI orbitals
may be different,  which may affect the nature of the induced
pairing potential~\cite{PhysRevB.99.161301}. In the supplementary material, we present the numerical results by considering that the signs and magnitudes of interlayer hopping constants are orbital dependent. Our results indicate that our main conclusions remain qualitatively the same even when the signs and magnitudes depend on the orbitals~\cite{supp}.
Therefore, for illustration,
we assume that the hopping constants are independent on the orbitals, with $t_{p\tau\tau^\prime}\equiv t_p$.
The other parameters are chosen as $t=1$, $m=5$, $A=0.5$, $t_1=-1$, $t_2=1.3$, $t_3=t_4=-0.85$, $V=1$, and $N_z=100$.
Then, Eq. (2) and Eq. (6) are effective models for describing the 3DTI and FeSC materials, respectively.
The energy unit $t$ is estimated to be approximately $0.2$ eV according to the first principles band calculation of FeSC materials~\cite{PhysRevLett.101.087004}.

\section{Results and Discussion}

  \begin{figure}
  	\centering
  	\includegraphics[width=3.3in]{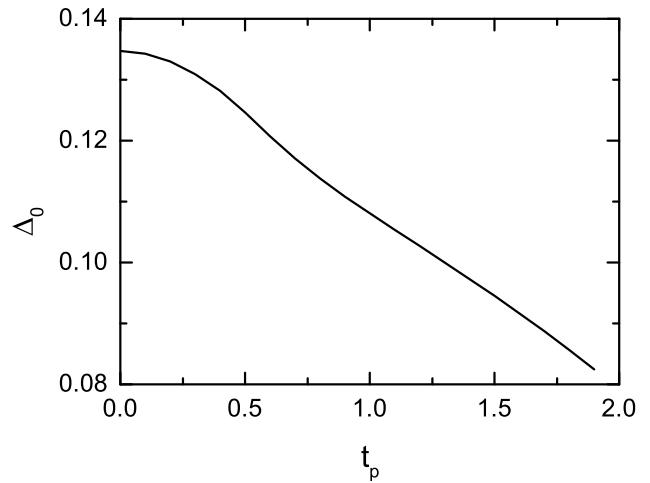}
  	\caption{Mean-field superconducting order parameter $\Delta_0$ as a function of the coupling strength $t_p$.
  	}\label{fig1}
  \end{figure}
  
  Before we present our theoretical results,
  we note the experimental realization of the 3DTI/FeSC heterostructure. Experimentally, this heterostructure was first realized by growing 3DTI films layer by layer on an FeSC substrate with the molecular-beam epitaxy technique~\cite{He2014,PhysRevB.97.224504,doi:10.1126/sciadv.aat1084}. With this heterostructure, one can directly measure the physical properties of the 3DTI layer through STM experiments and ARPES experiments. Additionally, successful epitaxial and
  contiguous growth of an FeSC material on a 3DTI material was recently reported~\cite{doi:10.1021/acs.nanolett.1c01703}. Moreover, very recently, various superlattices consisting of
  alternating 3DTI layers and FeSC layers were fabricated with the pulsed laser deposition technique~\cite{ZHANG2022163396}. With these heterostructures, the physical properties of the FeSC layer can also be directly measured. Therefore, on the theoretical side, investigations on both the 3DTI layer and the FeSC layer are insightful and may be used for comparison with later experiments.

  We first present the numerical result of the superconducting order parameter $\Delta_0$ as a function of the coupling constant $t_p$ in Fig.~1.
  Here, a kind of inverse proximity effect is revealed, namely,
  as $t_p$ increases, the pairing order parameter $\Delta_0$ monotonically decreases. At the mean-field level, the order parameter $\Delta_0$ is proportional to the superconducting transition temperature $T_c$. Our results indicate that the superconducting transition temperature of the 3DTI/FeSC coupled system is generally smaller than that of the pure FeSC material.
  Such suppression is due to hybridization of the energy bands.
  Actually, for a heterostructure including a superconducting material and a nonsuperconducting material, an effective pairing term will generally be induced in the nonsuperconducting material, known as the proximity effect. At the same time, unpaired quasiparticles will be induced back in the superconducting material through the inverse proximity effect. These quasiparticles generally suppress the superconductivity of the whole system. Here, the number of induced quasiparticles is expected to increase with the coupling strength of the system $(t_p)$; as a result, the superconducting order parameter $\Delta_0$ decreases as $t_p$ increases, as shown in Fig.~1.
  Previously, the suppression of the superconductivity in a heterostructure has also been reported experimentally and theoretically~\cite{Zareapour2012,PhysRevB.99.094505}.

  When the coupling strength is nonzero, an effective superconducting pairing term is expected to be induced in the surface of the 3DTI, known as the proximity effect. This effective pairing can be studied through the mean-field pairing order parameter from Eq. (11). The intensity plot of the mean-field pairing order parameter at the surface of the 3DTI as a function of the momentum ${\bf k}$ with $t_p=0.5$ is presented in Fig.~2(a). The contour plot of the zero value is plotted as dashed lines, indicating the nodal lines of the system. The order parameter is positive near the $X=(\pi,0)$ point and its symmetric points and negative near the $\Gamma=(0,0)$ and $M=(\pi,\pi)$ points. These properties are qualitatively consistent with the $s_{\pm}$ pairing symmetry of the FeSC material~\cite{RevModPhys.83.1589}. In contrast, significant differences between the induced pairing symmetry and the original $s_{\pm}$ pairing symmetry exist. First, although here, the effective pairing term exhibits positive and negative values, the positive magnitude is much smaller than the negative magnitude.
  Second, the induced order parameter has only twofold symmetry. The previous fourfold symmetry ($\mathcal{C}_4$ rotational symmetry) for the $s_{\pm}$ pairing is broken. This indicates that an additional $d$-wave component pairing term is induced.
  
  Previously, based on the microscopic model, broken $\mathcal{C}_4$ rotational symmetry was also indicated in a system including a two-dimensional topological insulator and a superconductor and discussed ~\cite{PhysRevB.81.241310,PhysRevB.103.024517}.
  This symmetry breaking is due to the spin-orbit coupling term of the topological insulator.
  The wavevector $(k_x,k_y)$ will change to $(-k_y,k_x)$ under the $\mathcal{C}_4$ operation.
  Due to the existence of the spin-orbit interaction [$A\neq 0$ in Eq. (2)], the 3DTI Hamiltonian varies under the $\mathcal{C}_4$ operation, namely, $H_{3DTI}(k_x,k_y)\neq H_{3DTI}(-k_y,k_x)$. Since the eigenvalues of $H_{3DTI}(k_x,k_y)$ and $H_{3DTI}(-k_y,k_x)$
  are identical, the energy bands of the pure 3DTI system still have $\mathcal{C}_4$ symmetry. However, as previously discussed, when the 3DTI couples to another system,
  generally, the $\mathcal{C}_4$ rotational symmetry will be broken by the proximity effect~\cite{PhysRevB.81.241310,PhysRevB.103.024517}.

  \begin{figure}
  	\centering
  	\includegraphics[width=3in]{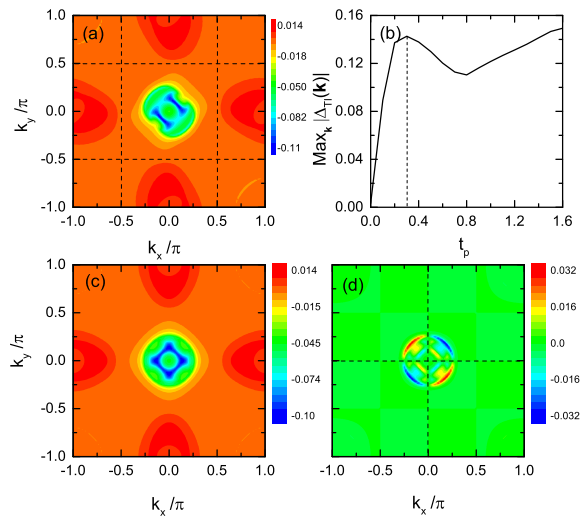}
  	\caption{(Color online) Numerical results of the proximity-induced pairing order parameter at the surface of the 3DTI. (a) Intensity plot of the momentum-dependent pairing order parameter. The dashed lines indicate the nodal lines. (b) Maximum pairing order parameter as a function of the coupling strength $t_p$.
  		(c) $s$-wave component of the pairing
  		order parameter. (d) $d$-wave component of the pairing
  		order parameter.
  	}
  \end{figure}

  We plot the magnitude of the induced pairing order parameter as a function of the coupling strength in Fig.~2(b). Quite interestingly, here, the order parameter magnitude is nonmonotonic versus the coupling strength. It reaches a local maximum value at approximately $t_p=0.3t$ (approximately 0.06 eV). Experimentally, for a topological superconducting material, a large pairing gap is important for identifying the possible zero modes. Intuitively, one may take for granted that a larger coupling strength is better for realizing an effective topological superconductor with a larger pairing gap, while our calculation indicates that a relatively small coupling strength may also induce a larger pairing gap. There exists an optimal coupling strength where the induced pairing order reaches a local maximum value.
  This result is of interest and may be applicable for realizing an effective topological superconductor with a heterostructure.
  
  The twofold symmetry of the order parameter shown in Fig.~2(a) indicates the existence of a $d$-wave component. We separate the whole pairing order parameter $\Delta_{TI}({\bf k})$ into the $s$-wave component $\Delta_{TIs}({\bf k})$ and the $d$-wave component $\Delta_{TId}({\bf k})$, with
  $\Delta_{TI}({\bf k})=\Delta_{TIs}({\bf k})+\Delta_{TId}({\bf k})$. $\Delta_{TIs}({\bf k})$ and $\Delta_{TId}({\bf k})$
  are expressed as $\Delta_{TIs}({\bf k})=1/2[\Delta_{TI}(k_x,k_y)+\Delta_{TI}(k_y,-k_x)]$ and $\Delta_{TId}({\bf k})=1/2[\Delta_{TI}(k_x,k_y)-\Delta_{TI}(k_y,-k_x)]$. The intensity plots of $\Delta_{TIs}({\bf k})$ and $\Delta_{TId}({\bf k})$ are plotted in Figs.~2(c) and 2(d), respectively. A small $d$-wave component around the $\Gamma$ point is revealed, with additional nodal lines $k_x=0$ and $k_y=0$ existing in this component. Therefore, our results indicate that the proximity-induced pairing order parameter symmetry is not necessarily identical to the previous order parameter symmetry of the superconductor. Similar conclusions have also been obtained from previous theoretical and experimental results~\cite{wang2013fully,doi:10.1126/sciadv.aat1084,PhysRevB.91.235143,PhysRevB.93.035140,PhysRevB.96.064518}. Moreover, as verified in Ref.~\cite{PhysRevLett.100.096407}, when an $s$-wave pairing term is induced in the 3DTI surface, the system is formally equivalent to a spinless
  $p+ip$ superconductor, providing an effective platform to realize Majorana zero modes.

  Considering both the surfaces of the 3DTI ($z=1$ and $z=N_z$) coupled with an FeSC layer,
  the energy bands can be obtained by diagonalizing the whole Hamiltonian [Eq. (1)].
  We plot the energy bands of the whole system along the $k_y=0$ direction with $t_p=0$ and $t_p=0.5$ in Figs.~3(a) and 3(b), respectively. Without the coupling term ($t_p=0$), as shown in Fig.~3(a), the energy bands of 3DTI and FeSC are independent. The energy bands of the 3DTI system include fully gapped bulk states and gapless surface states crossing the Fermi energy at the $\Gamma$ point~\cite{supp}. The energy bands of the FeSC material are fully gapped. In the presence of the coupling term with $t_p=0.5$, an obvious energy gap (approximately 0.1) is opened, as shown in Fig.~3(b). The gap magnitude is consistent with the order parameter magnitude shown in Fig.~2(b).

  \begin{figure}
  	\centering
  	\includegraphics[width=3in]{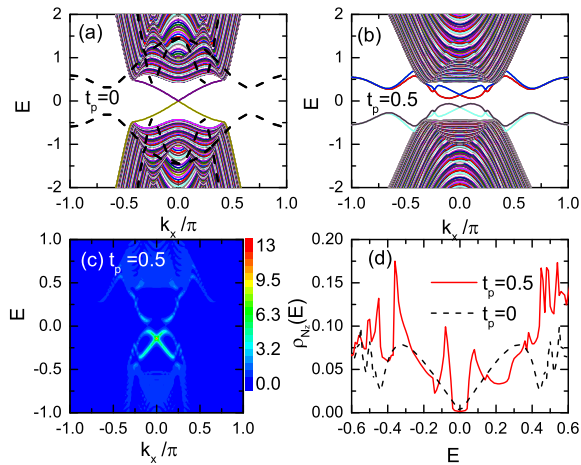}
  	\caption{(Color online) (a) Eigenvalues of the Hamiltonian with $k_y=0$ and $t_p=0$. (b) Eigenvalues of the Hamiltonian with $k_y=0$ and $t_p=0.5$.
  		(c) Spectral function at the $z = N_z$ surface with $k_y=0$ and $t_p=0.5$. (d) LDOS with $t_p=0$ and $t_p=0.5$.
  	}
  \end{figure}

  Experimentally, the proximity-induced energy gap may be investigated through the ARPES or STM technique. Theoretically, the ARPES and STM results are described by the spectral function and the LDOS, respectively. We plot the spectral function as a function of the energy $E$ and the momentum $k_x$ at the 3DTI surface with $k_y=0$ and $t_p=0.5$ in Fig.~3(c). The corresponding LDOS at the system surface without and with the coupling term is plotted in Fig.~3(d).
  The surface energy bands are revealed by the spectral function. The spectral function is zero at low energies of approximately $|E|<0.1$, indicating fully gapped behavior. The surface spectral function is qualitatively consistent with the energy bands presented in Fig.~3(b). Additionally, the proximity-induced superconducting gap can be clearly revealed through the LDOS. As $t_p=0$, the system is a topological insulator. Due to the existence of the gapless surface state, the surface LDOS has a large V-shaped energy gap. As $t_p$ increases to $0.5$, the spectrum becomes U-shaped, with the intensity being zero at low energies. Moreover, two coherent peaks at energies of approximately $\pm0.1$ are clearly observed. This indicates that an effective superconducting gap with an energy of approximately 0.1 is induced at the surface of the 3DTI. The U-shaped gap indicates that the system is fully gapped, consistent with the numerical results of the spectral function. Our numerical results indicate that the proximity effect can be studied and identified through ARPES and STM experiments~\cite{RevModPhys.75.473,RevModPhys.79.353}.

  \begin{figure}
  	\centering
  	\includegraphics[width=3in]{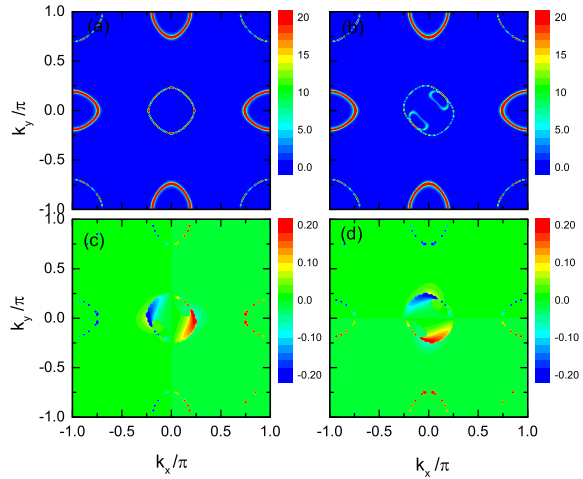}
  	\caption{(Color online) Numerical results for the FeSC layer in the normal state with $\Delta_0=0$. (a) Zero-energy spectral function with $t_p=0$. (b) Zero-energy spectral function with $t_p=0.5$. (c) Real part of the spin-flipped mean-field order parameter [Re$\lambda({\bf k})$]. (d) Imaginary part of the spin-flipped mean-field order parameter [Im$\lambda({\bf k})$].
  	}
  \end{figure}
  
  Now let us discuss in more detail the inverse proximity effect and investigate how the coupling affects the FeSC material. We first study the physical properties of the FeSC layer in the normal state. The intensity plots of the zero-energy spectral function in momentum space without and with the coupling are presented in Figs.~4(a) and 4(b), respectively. As is known, the zero-energy spectral function should be maximum at the Fermi momentum. Therefore, the normal-state Fermi surface can be obtained from Figs.~4(a) and 4(b). As $t_p=0$, the normal-state Fermi surface of the FeSC material has fourfold symmetry. It includes hole pockets surrounding the $\Gamma$ and $M$ points and electron pockets surrounding the $X$ point and its symmetric points. As $t_p$ increases, the fourfold symmetry is broken. The previous $\Gamma$ pocket is distorted. Two additional small pockets emerge inside the $\Gamma$ pocket. These additional pockets are due to the band mixing effect and may be detected by experiments. This may be used to judge whether the 3DTI material and the FeSC system are coupled in the heterostructure. Moreover, the normal-state Fermi surface in an unconventional superconductor is generally important and may determine many physical quantities. Here, the evolution of the normal-state Fermi surface with increasing coupling strength is of interest and is worthy of further study.

  In the original Hamiltonian describing the FeSC material, there are no spin-orbit coupling terms. However, here, due to the coupling with the 3DTI system, an effective spin-orbit coupling term may be induced in the FeSC layer, which can be described by a spin-flipped mean-field order parameter [Eq. (15)].
  The real and imaginary parts of the spin-flipped mean-field order parameter in the FeSC layer [Re$\lambda({\bf k})$ and Im$\lambda({\bf k})$] are displayed in Figs.~4(c) and 4(d), respectively.
  As seen, a spin-flipped term is indeed induced. Here, both the real part and the imaginary part of the order parameter have odd parity. The complex order parameter appears to have the form $\sin k_x-i\sin k_y$, qualitatively consistent with a general spin-orbit coupling term. Our results indicate that an effective spin-orbit coupling term is indeed proximity induced in the FeSC layer and may affect the topological properties of the system.
  
  \begin{figure}
  	\centering
  	\includegraphics[width=3in]{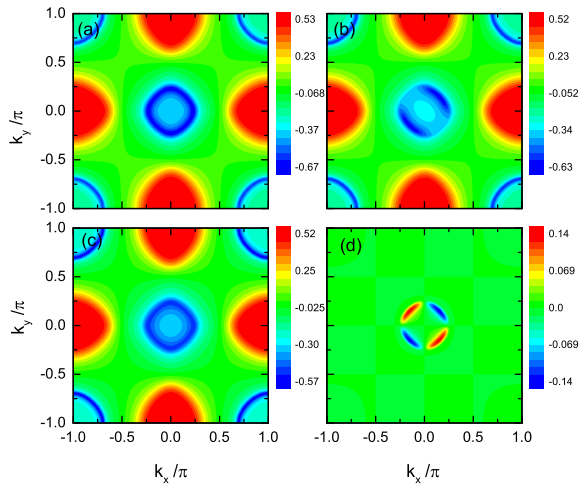}
  	\caption{(Color online) Numerical results of the pairing order parameter in the FeSC layer.
  		(a) Intensity plot of the momentum-dependent pairing order parameter with $t_p=0$. (b) Intensity plot of the momentum-dependent pairing order parameter with $t_p=0.5$. (c) $s$-wave component of the order parameter with $t_p=0.5$. (d) $d$-wave component of the order parameter with $t_p=0.5$. }
  \end{figure}
  
  We turn to discussing how the coupling affects the superconducting state. The mean-field pairing order parameters in the FeSC layer without and with interlayer coupling are presented in Figs.~5(a) and 5(b), respectively.
  First, as indicated in Fig.~1, the order parameter $\Delta_0$ is suppressed due to the interlayer coupling.
  Here, as shown in Figs.~5(a) and 5(b), the mean-field order parameters in the FeSC layer are indeed suppressed,
  but the fact that such suppression mainly occurs near the $\Gamma$ pocket should be emphasized. The energy gaps around the other Fermi pockets are nearly unaffected. Second, we expect that the pairing symmetry may be affected by the interlayer coupling.
  Similar to the 3DTI layer, here, in the presence of the interlayer coupling, the fourfold symmetry is also broken [Fig.~5(b)]. We also separate the whole order parameter in Fig.~5(b) into the $s$-wave component and the $d$-wave component. The two components are displayed in Figs.~5(c) and 5(d), respectively. The $s$-wave component of the order parameter is qualitatively the same as that of the pure FeSC material shown in Fig.~5(a), while in the presence of coupling, an additional $d$-wave component is induced, with $k_x=0$ and $k_y=0$ being the nodal lines.
  Such a symmetry breaking effect may also be detected by experiments and used to identify the coupling of different systems.

  We now discuss the possible limitations and outlook of our present work. First, we employ a minimal two-band model to describe the FeSC material. This model considers only the $d_{yz}$ and $d_{xz}$ orbitals of the Fe ions. Moreover, the possible intrinsic spin-orbit coupling that may exist in some FeSC materials is also not considered.
  We expect that the low-energy physics of the FeSC material can be qualitatively described based on this model, while a more accurate model may be more realistic. Second, here, the Hamiltonian for the FeSC material includes only a single layer, neglecting interlayer coupling. In particular, the inverse proximity effect may be affected by the interlayer coupling of the FeSC material. Third, the coupling between the FeSC material and the 3DTI material is simplified as single particle hopping with the hopping constants being independent of the orbitals. The results may be different if a more complicated coupling term is employed.
  Finally, based on phenomenological theory and starting from an effective ${\bf k}\cdot {\bf p}$ model describing the edge states of the 3DTI, the two-dimensional surface of the 3DTI material was previously verified to be equivalent to a $p+ip$ superconductor when the pairing term is proximity induced.
  However, the topological description and the definition of the topological invariant based on the present microscopic model still require further study.

  Finally, we stress that our main results can be well understood by analyzing the Fermi surfaces of the original FeSC and 3DTI systems. Generally, we expect that the proximity effect should be strengthened when the Fermi surfaces of the two systems match. The Fermi surface of the FeSC material is presented in Fig.~4(a), with Fermi pockets around the $\Gamma$, $M$, and $X$ points.
  The 3DTI surface state crosses the Fermi energy at the $\Gamma$ point. Therefore, when the two systems are coupled, both the proximity effect and the inverse proximity effect around the $\Gamma$ Fermi pocket should be dominant. As a result, in the 3DTI layer, the negative pairing magnitude near the $\Gamma$ pocket is larger. The induced $d$-wave component is around the $\Gamma$ Fermi pocket. In the FeSC layer, the $\Gamma$ Fermi pocket is distorted. Additional small Fermi pockets around the $\Gamma$ pocket are induced. The effective spin-orbit coupling is also larger around the $\Gamma$ pocket. In the superconducting state, an additional $d$-wave pairing component also emerges near the $\Gamma$ pocket. Actually, here, the normal-state Fermi surface is important in determining the main results. Therefore, we can reasonably conclude that our main results should remain qualitatively the same even if another effective model for the FeSC material (with qualitatively the same Fermi surface) is considered.

\section{summary}

In summary, we study the topological insulator/iron-based superconductor heterostructure with a microscopic method. Based on the self-consistent calculation, our numerical results indicate that the superconducting order parameter decreases when the coupling between the topological insulator and the iron-based superconductor increases. Similar to the $s_{\pm}$ pairing symmetry of the iron-based superconductor, here, the proximity-induced pairing order parameter also exhibits negative and positive values. However, the negative order parameter at the $\Gamma$ pocket is significantly stronger due to the gapless surface state at the $\Gamma$ point of the topological insulator.
The order parameter in momentum space has only twofold symmetry. It can be separated into an $s$-wave component and a $d$-wave component. Its maximum magnitude does not monotonically increase with the coupling strength. An optimal coupling strength exists at which the order parameter reaches the maximum value
(approximately $t_p=0.3\approx 0.06$ eV). The proximity-induced pairing order parameter is theoretically studied through the spectral function and the LDOS. The inverse proximity effect is also studied. An effective spin-orbit effect is induced in the iron-based superconducting layer. The fourfold symmetry in the superconducting layer is also broken. Additional small pockets are induced in the normal-state Fermi surface. In the superconducting state, an additional $d$-wave component is identified. The main properties can be understood by exploring the Fermi surfaces of the two original systems.

\begin{acknowledgments}
	This work was supported by the NSFC (Grant No. 12074130), the Natural Science Foundation of Guangdong Province (Grant No. 2021A1515012340), and the Science and Technology Program of Guangzhou (Grant No. 202102080434).
\end{acknowledgments}

%

\end{document}